\documentclass[preprint2]{aastex}

\newcommand\OII{[O\,{\sc ii}]}

\newcommand\SII{[S\,{\sc ii}]}

\newcommand\NII{[N\,{\sc ii}]}

\begin{document}

\title{Abell 2255: Increased Star Formation and AGN Activity in a Cluster-Cluster Merger}
\author{Neal A. Miller\altaffilmark{1}} 
\affil{NASA Goddard Space Flight Center \\ UV/Optical Branch, Code 681\\Greenbelt, MD \ 20771}
\email{nmiller@stis.gsfc.nasa.gov}

\and

\author{Frazer N. Owen\altaffilmark{1}}
\affil{National Radio Astronomy Observatory\altaffilmark{2}, P.O. Box O, \\ Socorro, New Mexico 87801}
\email{fowen@aoc.nrao.edu}

\altaffiltext{1}{Visiting Astronomer, Kitt Peak National Observatory, National Optical Astronomy Observatories, which is operated by the Association of Universities for Research in Astronomy, Inc. (AURA) under cooperative agreement with the National Science Foundation.}

\altaffiltext{2}{The National Radio Astronomy Observatory is a facility of the National Science Foundation operated under cooperative agreement by Associated Universities, Inc.}

\begin{abstract} 
Deep VLA 1.4 GHz radio continuum imaging of Abell 2255 is presented. This cluster is among the better nearby candidates for rich cluster-cluster merger systems, with evidence including an elongated X-ray morphology, the presence of a radio halo, and substructure present in its galaxy distribution. Our radio observations reach an rms sensitivity of $\sim40\mu$Jy beam$^{-1}$, enabling us to detect (at 5$\sigma$) star formation rates as low as 1.4 M$_\odot$ year$^{-1}$ from the center of the cluster out to a radial distance of 3$h_{75}^{-1}$ Mpc. The radio data are complemented by optical imaging and a large spectroscopic database, allowing us to separate all galaxies with $M_{R_c}\leq-20$ into cluster members and foreground/background galaxies. The spectra are also used to associate the galaxies' radio emission with either star formation or AGN. 

We compare the resulting cluster radio galaxy population with those of nineteen other nearby Abell clusters, and find strong evidence for an increase in the frequency of radio galaxies in Abell 2255. This increase is seen in two separate types of galaxies: powerful radio AGN and optically-faint star forming galaxies. The optical spectra of the latter often indicate current or recent starbursts, and these galaxies appear to be distributed along an axis perpendicular to the probable merger axis. We assess these factors in light of models of galaxy evolution, and suggest that the cluster-cluster merger is responsible for triggering galaxy activity in Abell 2255.
\end{abstract}
\keywords{galaxies: clusters: general --- galaxies: clusters: individual (Abell 2255) --- galaxies: evolution --- galaxies: radio continuum}

\section{Introduction}

In hierarchical structure scenarios, clusters of galaxies grow by the accretion of outlying galaxies and groups. In some cases, bonafide clusters themselves can merge. These cluster-cluster mergers are extremely energetic, involving kinetic energies up to $10^{64}$ ergs. On cluster-wide scales, the effect of this dumping of energy is apparent across a broad range in wavelength. The X-ray emission of cluster-cluster mergers is often elongated and can display discrete substructures and temperature variations \citep[e.g.,][]{mark2000,vikh2001}. At radio wavelengths, low surface brightness halos and relics \citep[e.g.,][]{jaff1979,hani1982,giov1999} seem to be associated with mergers as shocks can accelerate (or re-accelerate) electrons and ions and thereby lead to synchrotron emission \citep[e.g.,][]{trib1993,fere1996,giov2000,buot2001,enss2001}.

Despite clear evidence of activity over the scale of clusters, the effect of cluster-cluster mergers on individual galaxies is less certain. Mergers present complex environmental factors, including large galaxy-galaxy velocity dispersions and bulk flows of intracluster gas. What is the net effect of such factors on the evolution of member galaxies? Various models have investigated this question, but the large range in environmental parameters produces differing conclusions. \citet{bekk1999} evaluated the time-varying tidal effects on a disk galaxy whose parent group was being accreted by a cluster. His simulations indicated that cluster-group mergers could lead to star formation bursts in member galaxies through transfer of large amounts of gas from the galaxies' disks to their nuclei. While the large relative velocities of galaxies in cluster-cluster mergers strongly inhibits direct galaxy-galaxy mergers, the cumulative effect of numerous high-speed passages of galaxies (``harrassment'') can also efficiently transfer gas to the galaxies' nuclei and lead to enhanced star formation \citep{moor1996,moor1998,fuji1998}. However, these studies generally ignore the effect of the intracluster gas on member galaxies. Pressure from this gas leads to competing effects: compression of the ISM in the moving galaxies can enhance star formation \citep[e.g.,][]{evra1991,roet1996}, while removal of gas due to ram pressure stripping can squelch it. \citet{fuji1999} argued through simulations that the net effect of this was to reduce star formation in member galaxies during cluster-cluster mergers, although they did not specifically address the effect of shocks in the ICM.

\citet{dwar1999} and \citet{owen1999} observed a complementary pair of clusters to investigate the Butcher-Oemler effect. These clusters were nearly identical in terms of their redshift and richness, yet radio and optical data showed that one possessed a high fraction of active galaxies while the other did not. The apparent cause was the dynamical state of the clusters, with the cluster containing more active galaxies being a strong merger candidate and the other cluster being more virialized. Similarly, further observational clues may be hidden in studies of clusters at higher redshift. The MORPHS collaboration \citep{dres1999} have studied ten intermediate redshift clusters to investigate galaxy evolution. They find evidence for increased activity in these clusters in the form of larger numbers of star-forming and post-starburst galaxies. However, the CNOC collaboration \citep[see][and subsequent papers]{yee1996} have studied clusters in a similar redshift range and seen significantly less evidence for such an effect. The primary difference in their studies is in how the samples were constructed, with the MORPHS clusters being identified optically and the CNOC clusters through X-ray luminosity. This biases the CNOC sample to more relaxed, virialized clusters, which could very well explain the differences in their member galaxy populations \citep{elli2001}. The increased activity seen by the MORPHS collaboration could be the result of their clusters more frequently representing merger systems.

In this paper, we assess the radio galaxy population of one of the better nearby examples of a merger between two clusters of galaxies. A compelling body of evidence supporting this scenario has been compiled for Abell 2255, a Northern hemisphere cluster at $z\sim0.08$. The cluster shows an elongated X-ray distribution whose peak is offset from the optical center of the cluster \citep[e.g.,][]{burn1995,fere1997}, and substructure in X-ray temperature maps of its intracluster gas \citep{davi1998}. Optically, the brighter galaxies are arranged in a chain, with the two brightest galaxies at the center of the cluster separated by nearly $3000$ km s$^{-1}$. This led to the hypothesis that Abell 2255 was actually two separate clusters seen in projection \citep{tare1976}, a result which appeared feasible given the small number of measured galaxy velocities at that time. Additional velocity measurements argued that Abell 2255 was, in fact, a bound system with an unusually high velocity dispersion \citep{stau1979,zabl1990,burn1995}. Even so, the total number of cluster galaxy velocities was modest and although no substructure was detected from these velocities the lack of data prohibited strongly dismissing its presence. Recently, an enormous increase in the number of measured velocities has become available \citep{hill2002,sdss2002}. These have enabled the detection of significant evidence for substructure and argue that Abell 2255 is a dynamically-active cluster being assembled from numerous groups \citep{hill2002}. Lastly, radio observations of the cluster also note an unusual level of activity. Several studies have noted a large number of head-tail radio galaxies in the cluster, as well as a diffuse radio halo \citep{jaff1979,harr1980,burn1995,fere1997}.

We have performed wide-field, high sensitivity radio observations of Abell 2255 with the National Radio Astronomy Observatory's (NRAO) Very Large Array (VLA). These observations encompassed the entire cluster, from its core out past the Abell radius ($r= 0 - 3$ $h_{75}^{-1}$Mpc), and identified galaxies whose 1.4 GHz radio emission amounted to $\sim3 \times 10^{21}$ W Hz$^{-1}$ and greater. This corresponds to all AGN and galaxies forming stars at rates in excess of 1.8 M$_\sun$ yr$^{-1}$ \citep[using the relation found in][]{yun2001}. The radio data are complemented by new wide-field optical images collected in $R_c$, multiband photometry from the Sloan Digital Sky Survey Early Data Release \citep[SDSS EDR;][]{sdss2002}, and the spectroscopic databases of \citet{hill2002} and the SDSS EDR. Consequently, we are able to explore the effect of the cluster merger on galaxies down to a very low level of activity. The results for Abell 2255 are then compared against a large sample of nearby clusters \citep{mill2002}, demonstrating the unusual properties of the cluster. The results are discussed in light of various models for galaxy evolution in cluster environments.

The paper is organized as follows. In Section \ref{sec:obs}, we discuss the data. This includes details about the observations and reductions for both the radio and optical data, and a discussion of the optical spectroscopy data which were used. The results are presented in Section \ref{sec:results}, including an overview of the cluster radio galaxy population, an analysis of how the radio galaxies are distributed, and statistical comparisons of the radio galaxy population of Abell 2255 relative to other nearby clusters. These findings are discussed in light of cluster evolutionary models in Section \ref{sec:discuss}, and final conclusions are presented in Section \ref{sec:conclude}. We have adopted $H_o = 75$ km s$^{-1}$ Mpc$^{-1}$ and $q_o=0.1$, meaning 1\arcsec{} corresponds to 1.28 kpc at Abell 2255.

\section{Observations and Reductions}\label{sec:obs}

\subsection{Radio}\label{sec:radio_obs}

Radio observations at 1.4 GHz were made with the VLA in two configurations. The higher resolution observations were performed during July 1998 while the VLA was in its B configuration, and lower resolution observations were performed during May 1999 in the D configuration. The B configuration provides a resolution of $\sim$5\arcsec{} but is insensitive to structures larger than $\sim$120\arcsec. Abell 2255 is known to possess a number of extended radio galaxies as well as diffuse emission unassociated with individual galaxies \citep{harr1980,burn1995,fere1997}, necessitating the inclusion of the lower resolution D array data.

The observational strategy paralleled that of other wide-field radio surveys \citep[e.g., the NVSS and FIRST;][]{cond1998,beck1995}. A grid of 25 pointings was used to cover the cluster from its core out to large radii ($\gtrsim$40\arcmin, or about 3 Mpc). These pointings were arranged in a hexagonal pattern, with the primary 19 pointings having a grid spacing of 16\arcmin. The remaining 6 pointings formed a smaller hexagon surrounding the central pointing of the main grid. This configuration was adopted to provide near-uniform sensitivity across the cluster, with extra detail in the cluster core. For an excellent overview of the technique, from observation through to construction of the final wide-field image, the reader is directed to \citet{cond1998}.

Both B and D array observations were made in ``line mode'' to greatly reduce bandwidth smearing. This is also required in order to properly ``clean'' sources away from the pointing center, thereby reducing the noise caused by confusion. Thus, seven channels of 3.125MHz bandwidth were used for each intermediate frequency and polarization. The integration time for each pointing was 17 minutes while in the B configuration, and 8.5 minutes in D. The pointings were made at the same hour angle for both B and D to facilitate combining the uv data and to correct for the ``3D effect.''

Data reduction was performed using NRAO's Astronomical Image Processing System (AIPS). Flux calibration was performed via observation of 3C286, and the flux density scale should be accurate to $\lesssim2\%$ \citep{baar1977}. The flux density scale was further confirmed via comparison with NVSS fluxes for unresolved sources. Calibration of both the individual bandpasses and the phases was provided by observations of 3C343, performed roughly every hour. This source is located approximately 4$^\circ$ from the center of Abell 2255. 

The pointings were each reduced individually using the AIPS task IMAGR. The first step involved determining the optimal choices of parameters to satisfy the dual considerations of minimizing noise and insuring consistent beam shapes across all pointings. Three parameters were varied, corresponding to the weighting of the uv data (ranging from uniform to natural), smoothing of the uv data, and applying various Gaussian tapers to the uv data. An image for each pointing was then created after applying a strong taper to the data (i.e., devaluing the longer baselines to create a lower resolution map extending well past the first null in the primary beam). This image was used to identify flanking sources falling in the beam's sidelobes, which is necessary to properly remove their contributions in the primary beam. Once identified, coordinates for the flanking sources were obtained from the NVSS, which has properly treated the shift in apparent position due to the 3D effect. 

Producing the final maps for each pointing was then performed iteratively. The B-array data were reduced first, followed by the D-array data. The primary beam was imaged in four separate facets to reduce the 3D effect, with flanking sources boxed and cleaned. The resulting maps were inspected and all sources boxed, followed by a second imaging of the uv data. The output of this run was then used to self-calibrate the uv data (phases only), with the new self-calibrated data then being used in another imaging run. The output of this run was then used to further self-calibrate the data (now on both amplitude and phase), and the resulting uv data was inspected and interference excised. Any new faint sources which had appeared in the images were boxed, and a semi-final imaging of the modified uv data was performed. The net result of these steps was typically a reduction of $\sim15\%$ in the noise relative to the initial mapping. Once this point had been reached for both the B and D array datasets, the two were combined into a single uv dataset and imaged. A circular restoring beam of 5.9\arcsec{} was used to create the final images for each pointing. 

The reduced images for each pointing were then stitched together to form the final mosaic maps. This was performed by weighting each location in a given pointing by its S/N, which is just the power pattern of the primary beam. In this step, each map was truncated at $70\%$ of the primary beam (about 20\arcmin). Thus, locations at the pointing centers receive full weight in the final output map whereas locations distant from the pointing centers are devalued. The final mosaic had an rms noise of about 30 $\mu$Jy beam$^{-1}$ in the center. At the very edges, where fewer individual pointings contribute to the mosaic, the noise level was still only $\sim45$ $\mu$Jy beam$^{-1}$. Our radio maps do not successfully image the extended diffuse radio halo. Its presence is likely indicated by a slight increase in the rms noise in the regions of the map where it should be located (from around 30 $\mu$Jy to just under 40 $\mu$Jy).

The AIPS task SAD (``search and destroy'') was used to identify sources in the final mosaic. This task was directed to locate all sources with peak fluxes greater than four times the rms noise level determined over the surrounding $\sim30$\arcmin. SAD fits these sources with Gaussians and notes their positions and fluxes. It also creates a residual map, which was inspected to locate any extended sources not well fit by Gaussians. These were manually added to the lists, with fluxes determined directly by boxing the sources. Each radio source was then inspected, and only those with either peak or integral fluxes greater than five times the local noise (over the surrounding 5\arcmin) were included in the final radio source lists. In total, the radio source list contained over 1200 members.

\subsection{Optical Imaging}\label{sec:opt_image}

Optical images ($R_c$) of Abell 2255 were obtained during July 1999 using the KPNO 0.9 meter telescope. These were taken with the Mosaic camera, providing a 59\arcmin{} field at a pixel scale of 0.43\arcsec. To cover the spacings between the eight individual chips in Mosaic, we adopted the standard dither sequence of five images per telescope pointing. The wide field of Mosaic allowed Abell 2255 to be imaged well past its Abell radius in only four pointings. A fifth, deeper pointing was centered on the cluster core. The net exposure time was 750 seconds for each of the four shallower images and 1500 seconds for the deeper central pointing. Observations of three Landolt Selected Areas \citep{land1992} were performed at a range of airmass (1.2 to 2.3) to provide photometric calibration.

The reductions were done following the steps suggested in the NOAO CCD Mosaic Imager User Manual\footnote{Jacoby et al., \url{http://www.noao.edu/kpno/mosaic/manual/} and \url{http://iraf.noao.edu/projects/ccdmosaic/Reductions}}. A sequence of eight bias frames were stacked to create a superbias, and a flat field to handle pixel-to-pixel sensitivity variations in the CCDs was created from a set of eight dome flats. The wide field of Mosaic presents an additional challenge in flat fielding. To produce a flat which accurately corrects for the illumination pattern of the telescope and detector, all of the night's target exposures were used (including observations of standards and another cluster, Abell 2256). These were scaled by their modes and stacked after careful rejection of peaks (cosmic rays, stars, and galaxies) to create a sky flat. This sky flat was then boxcar smoothed to approximately the same signal-to-noise as the dome flat. The CCDPROC task in the MSCRED package was then used to overscan subtract, bias subtract, flat field, and trim the raw images using the aforementioned calibration frames. The five dithers for an individual pointing were then combined, with relative astrometry determined using $\sim$50 stars per field. The rms coordinate shifts among the dither images were less than 0.3 pixels. The stars also were used to match the intensity scales of the five dithered images, which were consistent to well under $1\%$ as would be expected for photometric conditions.

The three Landolt Selected Areas provided a total of $\sim70$ standard stars for photometric calibration. These were used to fit the dependence of magnitude on airmass, but no color term was applied. The derived relationship was $m = m_{inst} - 0.10X + 2.5\log t + C$, where $m_{inst}$ is the instrumental magnitude, $X$ is the airmass, $t$ is the time, and $C$ a constant. Comparison of derived magnitudes for the standard stars vs. their published values suggests the photometry is accurate to within 0.05 magnitudes. 

Astrometric registration for the images was performed using the USNO A2.0 catalog \citep{mone2000}. The USNO A2.0 catalog is the optimal choice for the astrometry, as its coordinate system is tied to the radio reference frame using distant quasars. The catalog also includes faint stars, which are an improvement over the brighter HST Guide Star Catalog stars because they will not be saturated in the images. The use of saturated stars leads to errors in fitted centers and consequently less accurate registration. Several hundred USNO A2.0 stars were used per Mosaic field. The coordinate solution relative to these stars was determined, and any stars which produced large residuals relative to this fit were removed (usually around 10 stars per Mosaic field). Presumably, these stars correspond to those with large proper motions. The residuals of the final astrometric solutions were around 0.4\arcsec.

Lists of optical sources were generated using the SExtractor software package \citep{bert1996}. Our magnitudes were derived using the Gunn-Oke \citep{gunn1975} metric aperture, which corresponds to a radius of 13.1 kpc using our adopted cosmology. Optical galaxies brighter than $M^*_R=-20$ \citep[i.e., $m^*_R+2$ using $M^*_R=-22$;][]{owen1989} were included in the formal list of potential cluster radio galaxies, as in \citet{mill2001}. This corresponds to $m_{R_c}=17.51$ for Abell 2255, adjusted for $A_{R_c}=0.07$. Finally, the images were visually inspected to confirm the star/galaxy segregations made by SExtractor.

The list of optical galaxies was then correlated with the list of radio sources. For this procedure, we adopt a maximum probability for chance coincidence of radio and optical sources of $0.5\%$ \citep[for a brief description, see][]{mill2001}. For Abell 2255, this amounted to a maximum positional separation of just over 6\arcsec. In practice, only one galaxy with unresolved radio emission had a radio-optical separation of more than 3\arcsec. Overlays of radio contours on the optical images were also used to locate radio galaxies with extended emission, whose fitted Gaussian positions need not necessarily correspond to optical positions. The resulting list contained 64 candidate radio galaxies brighter than $M_{R_c}=-20$ and within 3 Mpc of 17:12:45 +64:03:54, the adopted center coordinate of Abell 2255 \citep[corresponding to the centroid of the cluster X-ray emission;][]{fere1997}.

Abell 2255 is also within the regions included in the SDSS EDR. While we have used our own $R_c$ imaging to create our formal list of radio galaxy candidates, we have also obtained the SDSS $u'g'r'i'z'$ photometry for all galaxies in the surveyed area. 

\subsection{Optical Spectroscopy}

An aggressive campaign of optical spectroscopy has been staged for this cluster. \citet[][Paper 1]{hill2002} details multifiber spectrograph observations, and many of the radio galaxies were observed individually using long slits \citep[see][]{mill2001,mill2002}. Finally, the SDSS EDR contains hundreds of measured velocities in the Abell 2255 region. From the center of the cluster out to a radial separation of one degree, there are 323 cluster velocities presented in Paper 1.

Consequently, velocity measurements existed for all 64 of the candidate cluster radio galaxies. Adopting a systemic velocity of 23988 km s$^{-1}$, all galaxies within three times the cluster velocity dispersion of 1201 km s$^{-1}$ were taken to be cluster members \citep[values derived from Paper 1 data for our 3 Mpc limit, using the biweight estimators of][]{beer1990}. This produced a list of 46 cluster radio galaxies, along with four foreground radio galaxies and 14 background radio galaxies. Two of the apparent background radio galaxies had somewhat uncertain velocities. The spectra of these galaxies exhibited no emission lines, and cross correlation of their spectra with a variety of templates yielded their velocities at relatively low confidence \citep[as parametrized by $R$; see][]{tonr1979}. These velocities were adopted, the viability of which will be discussed in greater detail below.

In the absence of spectroscopy, galaxy colors may be used to produce a rough assessment of velocities. We have correlated the velocity database with the SDSS EDR photometry. In Figure \ref{fig-colors} we plot the $r'$ magnitudes vs. $g' - i'$ colors for all galaxies in the Abell 2255 region with measured velocities. We have used only galaxies with $r' < 18$ for this figure (slightly fainter than our $m_{R_c}\leq17.51$ limit, corrected for the different filter bandpasses). Galaxies within the cluster are represented by filled circles, while foreground and background galaxies are plotted as open triangles and open circles, respectively. Cluster radio galaxies (within the adopted magnitude and radial limits) are depicted as crosses. The cluster red sequence is obvious, with many galaxies clustered along the line defined by $g' - i' \approx 1.35$. This is the expected $g' - i'$ color for ellipticals and S0s at the cluster redshift \citep{fuku1995}. It may also be seen that background ellipticals follow a well-defined sequence extending up and to the right (i.e., redder colors with increasing apparent magnitude). The two apparent background galaxies with somewhat uncertain velocities are plotted as crosses. These galaxies lie along the sequence of background elliptical galaxies, and consequently we are confident that they are background sources. Finally, about half of the foreground galaxies belong to a group at $\sim16100$ km s$^{-1}$, including two which were also radio sources.

\placefigure{fig-colors}

\placetable{tbl-clusRG}

Table \ref{tbl-clusRG} presents the data for the confirmed radio galaxies. This includes positions, 1.4 GHz flux densities, photometry, and velocities for all 46 cluster radio galaxies. Additional radio sources are presented in Table \ref{tbl-moreRG}. These include the 18 foreground and background galaxies identified using the formal criteria for the sample, plus an additional 19 radio galaxies with velocity measurements. These latter sources were identified by correlating the complete velocity database with the radio source list, and include: 1) two cluster radio galaxies located just outside the 3 Mpc radial limit, 2) three cluster radio galaxies which were optically too faint for inclusion in the sample ($m_{R_c}>17.51$), 3) one foreground radio galaxy with $m_{R_c}>17.51$, and 4) 13 background radio galaxies with $m_{R_c}>17.51$. Included among the background sources are three SDSS QSOs with $z>1$ (J170911+642210, J171330+644253, and J171400+640940). Their radio fluxes are below the limit of the FIRST survey, which is being used to identify radio QSOs from SDSS data. In total, this study has ascribed radio emission to 83 galaxies with measured velocities.

\placetable{tbl-moreRG}

The cluster radio galaxies were classified on the basis of their emission and absorption features. In general, this classification amounted to identifying the source of the radio emission as due to star formation or an active nucleus. All galaxies with emission line spectra were classfied using line ratio diagnostics \citep[e.g.,][]{bald1981,veil1987} after correcting for stellar Balmer absorption. Details of the line measurement procedure and galaxy classification are described in \citet{mill2002}. The same procedure was applied to the SDSS data, except that the SDSS uses 3\arcsec{} fibers for collection of spectra and consequently provides only a nuclear spectrum. In addition to these classifications, we have classified the radio galaxies using the scheme adopted by the MORPHS collaboration \citep{dres1999}. The classifications for the cluster radio galaxies may be found in Table \ref{tbl-clusRG}.

\section{Results}\label{sec:results}
\subsection{Overview of the Radio Source Population}\label{sec:over}

As has been noted in prior studies, Abell 2255 has an unusually large number of galaxies with extended radio emission. There are six cluster radio galaxies with FR I or II morphologies, which ranks it among the richest of all Abell clusters in terms of number of such sources \citep[e.g., see ][]{ledl1995}. These powerful radio galaxies are not exclusively located in the core of the cluster. In fact, one of the noted cluster radio galaxies just outside the 3 Mpc radial limit is an FR I \citep[the extreme clustercentric distance of this radio galaxy was also noted in][]{harr1980}. The six cluster radio galaxies with FR I and FR II morphologies are shown in Figure \ref{fig-otrg}--\ref{fig-embryo}, along with the FR I associated with the cluster but outside the 3 Mpc limit (Figure \ref{fig-bean}). An eighth galaxy with extended radio emission (J171406+641602, shown in Figure \ref{fig-faker}) is clearly a background radio galaxy. \citet{fere1997} noted this source as a potential cluster member based on its radio morphology, but its measured velocity places it at $z\sim0.25$. It may also be seen to lie along the locus of background ellipticals in Figure \ref{fig-colors} ($r'=17.63$, $g' - i' = 2.15$). Several other extended radio sources with apparent optical hosts fainter than our cutoff magnitude were noted, although none of these have measured velocities and all are likely at higher redshift based on their optical magnitudes, colors, and radio morphologies and fluxes.

\placefigure{fig-otrg}
\placefigure{fig-double}
\placefigure{fig-goldfish}
\placefigure{fig-beaver}
\placefigure{fig-embryo}
\placefigure{fig-bean}
\placefigure{fig-faker}

The radio observations also detected numerous star forming galaxies in the cluster. In total, at least 26 galaxies exhibit emission lines indicative of current star formation. In several instances, these emission lines are very strong and narrow and indicate a starburst galaxy of type e(b) in the MORPHS classification system. Formally, we identify two of these type galaxies although we note that a third galaxy may also be of this type. The wavelength coverage of our spectroscopy of this galaxy narrowly missed the \OII{} line (rest frame 3727$\mbox{\AA}$). This galaxy had an H$\alpha$+\NII{} equivalent width $>140\mbox{\AA}$, and therefore probably has an \OII{} strength sufficiently large as to classify it as e(b). In addition to these three galaxies, we note that a few other galaxies have spectra resembling starbursts (e.g., J171223+640829) but do not quite meet the requirement to be e(b) galaxies and are therefore classified as e(c) in the MORPHS classification system. Such spectra are more representative of a continuous star formation history \citep{pogg1999}.

There is also a relatively large population of galaxies with e(a) spectra in the cluster. These galaxies have emission of \OII{} indicating current star formation, but also have unusually strong H$\delta$ absorption. \citet{pogg1999} used both models and observational arguments to demonstrate that galaxies with this type of spectrum are most likely associated with dusty starbursts. The dust extinction is age dependent, with younger stars more heavily obscured. We identify at least five such galaxies. While J171117+642033 has an H$\delta$ equivalent width of 4.8$\mbox{\AA}$ in absorption, the error in this quantity is very large (3.2$\mbox{\AA}$) and we tentatively classify it as an e(c) galaxy. There is also a k+a galaxy (i.e., a post-starburst or ``E+A'' spectrum) from which the H$\alpha$, \NII{}, and \SII{} lines indicate current star formation (J171334+634926). Galaxies with such spectra are manifestations of the e(a) class in which dust extinction has entirely removed the \OII{} emission, or reduced it below the noise level of the spectrum \citep{pogg1999}. 

Two further galaxies may be more extreme examples of star formation hidden by large dust extinction. J171207+640832 has a k+a spectrum and no emission line evidence for star formation, yet its radio luminosity ($5.8 \times 10^{21}$ W Hz$^{-1}$) places it in the portion of the radio luminosity function which is dominated by star forming galaxies. Consquently, it may represent an extreme example of current star formation in highly dust-obscured regions \citep{smai1999,mil2001b}. One further cluster radio galaxy may be another example of this type of spectrum, but its spectrum is too noisy to state this with confidence (J171400+640542).

The remaining cluster radio galaxies appear to be AGN of lower radio luminosity. There were 12 such objects, seven of which exhibited emission lines representative of Seyfert or LINER activity. The five remaining galaxies had absorption line spectra and were associated with fairly bright elliptical and S0 galaxies. These appear to be a lower luminosity extension of the more powerful radio AGN.

\subsection{Radio Galaxy Positions and Relation to X-ray Emission}

Abell 2255 has been the target of several X-ray observations \citep{burn1995,fere1997,davi1998}, making data available for comparison of the X-ray emission of the cluster to the distribution of the radio sources. In Figure \ref{fig-xray} we plot the ROSAT PSPC data for the cluster. The large circle indicates the 3 Mpc radial limit of this study, whose center is the approximate centroid of the X-ray emission. The X-ray emission is clearly elongated in the East-West direction (which is also evident in the distribution of the brightest galaxies in the cluster core), arguing that this is the primary axis of the cluster-cluster merger \citep[e.g.,][]{burn1995}. \citet{fere1997} also demonstrate that the radio halo aligns well with the cluster X-ray emission.

\placefigure{fig-xray}
\placefigure{fig-agndist}
\placefigure{fig-sfdist}

The locations of the radio sources are indicated in Figure \ref{fig-agndist} and \ref{fig-sfdist} (they were omitted from Figure \ref{fig-xray} for clarity). The locations of the AGN are shown in Figure \ref{fig-agndist}, with diamonds representing AGN with absorption-line spectra (which include all of the powerful radio galaxies with extended emission) and triangles the emission-line AGN. For reference, the radio galaxy nearest the X-ray centroid is J171251+640424, an emission-line AGN with a LINER spectrum. Its separation from the X-ray centroid is about 45\arcsec. The centroid is also bracketted by three of the FR I galaxies and the FR II galaxy (Figure \ref{fig-otrg}--\ref{fig-goldfish}). A general correspondance between the AGN and the X-ray emission can be seen, with the central core of the AGN distribution being elongated in the East-West direction.

The star forming galaxies (Figure \ref{fig-sfdist}) do not exhibit the same East-West alignment. In fact, they appear to be distributed preferentially along an axis running from the Southeast to the Northwest. This is more evident when one considers the spectral type of the star forming galaxies. In particular, the star forming galaxies with evidence for a current or recent {\it burst} of star formation (the MORPHS classes e(a), e(b), and the k+a galaxy which is forming stars) are all along this general axis. The location of the k+a galaxy without any optical emission lines is also along this axis. The visual interpretation of an alignment of the star forming galaxies was tested through the use of a Fourier Elongation Test \citep[see][]{pink1996}. When all the star forming galaxies were included, there was no evidence for a prefered axis in the galaxy distribution. However, removal of the six star forming galaxies at radii larger than 2 Mpc (all of which are e(c) galaxies) produced a result significant at $\sim99\%$ confidence. That is to say, the distribution of the star forming galaxies within 2 Mpc of the cluster center is more elongated than 986 out of 1000 Monte Carlo simulations of the data.

Possible additional evidence for this prefered axis for the star forming galaxies may be seen in the optical morphology of one of the identified starburst galaxies. A grey-scale image of this galaxy (J171320+640034) is shown in Figure \ref{fig-ebgal}. At least three ``streamers'' can be seen pointing away from the galaxy toward the SE. The radio emission from the galaxy peaks along the central streamer. The absolute magnitude of this galaxy is $M_{R_c}=-20.1$, making it one of the fainter galaxies in the study. Its radio luminosity is $6.3 \times 10^{21}$ W Hz$^{-1}$.

\placefigure{fig-ebgal}

\subsection{Comparison to Other Clusters}

In order to assess whether Abell 2255 is truly unusual in terms of its radio galaxy population, statistical tests were performed which compared the cluster to a composite sample of 19 other nearby clusters \citep[][]{mill2002}. The full 20 cluster sample consists of 18 clusters studied using NVSS data, plus Abell 2255 and Abell 2256. The radio observations for Abell 2256 were performed concurrently with the Abell 2255 program, and are discussed in a separate paper \citep{mill2003}. The statistical tests effectively evaluate differences in the radio luminosity function (RLF) of Abell 2255 as compared to the RLF of the other clusters. The typical number of radio galaxies per cluster is around 20, which often limits the significance at which conclusions may be drawn for a given cluster. However, this problem is largely avoided in the case of Abell 2255 due to its large number of radio galaxies.

First, we ask whether the shape of the RLF for Abell 2255 differs from that of the other clusters. This was tested through application of a KS test, and requires no information other than the radio luminosities of the galaxies in all twenty clusters. A sample consisting of all clusters other than the cluster being studied is created, and the null hypothesis is that the radio luminosities of the cluster in question were drawn at random from the complete sample. This hypothesis could not be rejected with any reasonable level of confidence for any cluster in the sample, including Abell 2255. Thus, despite the unusually high number of powerful radio galaxies in Abell 2255, there is no evidence that such sources are more numerous than would be expected given the rest of the radio population in the cluster.

However, this does not examine the normalization of the RLF --- the frequency with which galaxies in Abell 2255 are radio sources relative to galaxies in other clusters. Rephrased, is the fraction of radio galaxies to total cluster galaxies greater in Abell 2255 than in other clusters? This question is similar to that used in other studies of galaxy evolution, such as the Butcher-Oemler effect \citep{butc1984}. A statistically significant larger fraction of radio galaxies would be an indication of increased activity within Abell 2255 (a higher normalization for the Abell 2255 RLF), which may then be used to assess the potential environmental attributes of the cluster which give rise to such an effect.

To formally test the radio galaxy fractions, we used a chi-square statistic:
\begin{equation}
\chi^2 = \sum_{i=1}^k \sum_{j=1}^{2} \frac{(f_{ij}-e_{ij})^2}{e_{ij}}.
\end{equation}
Here, the index $k$ implies that up to $k$ different groups could be compared to evaluate whether any given one differed from the others. For comparison of clusters individually against the collective sample created by the other clusters, $k=2$. The index $j$ refers to whether a galaxy is detected or not detected in the radio, so $f_{ij}$ represent the actual number of radio and non-radio galaxies and $e_{ij}=n_i\hat{\theta}$ the number of expected radio and non-radio galaxies. The latter quantity is determined based on the number of galaxies in the cluster being tested ($n_i$) and the radio galaxy fraction derived from the collective sample ($\hat{\theta}$). The resulting statistic is distributed as a chi-square with $k-1$ degrees of freedom. One potential weakness of this test is that it can be adversely affected by small number statistics; small expected values ($e_{ij}$) produce artificially high significance. Consequently, we have dismissed any results that included an $e_{ij}<1$.

The number of cluster members for each cluster was estimated by counting all galaxies within 3 Mpc as a function of magnitude. This total number was corrected for background counts by assuming that the surface density of background galaxies is represented by $N = \mathcal{N}10^{0.6m}$, where $\mathcal{N}=1.26 \times 10^{-5}$ galaxies steradian$^{-1}$. This value of $\mathcal{N}$ was determined directly from the regions at the outskirts of the poorer clusters in the collective sample, and is consistent with that found in other studies \citep[e.g., $\mathcal{N}=1.43 \times 10^{-5}$ and $1.04 \times 10^{-5}$ galaxies steradian$^{-1}$ for][respectively]{peeb1973,yee1999}. Finally, since the 18 nearer clusters of the sample all used the NVSS as their source of radio data we have only included galaxies with radio luminosities above $6.9 \times 10^{21}$ W Hz$^{-1}$, which amounts to a flux of 3.4 mJy at the farthest cluster for which NVSS data were used (note that this correponds to about 580 $\mu$Jy in Abell 2255, significantly above our achieved flux limit). The galaxy counts are summarized in Table \ref{tbl-rgnums}.

\placetable{tbl-rgnums}

The tests were also performed separately for cluster galaxies within 3 Mpc (the limit of our surveys) and 2 Mpc (about an Abell radius). There was substantially more variation among the clusters when using the 3 Mpc radius (see Table \ref{tbl-rgnums}), as it accentuates the contribution due to the background correction. Consequently, we will focus on the results obtained for galaxies within 2 Mpc of the cluster centers.

Two clusters were found to have significantly different radio galaxy fractions than the rest of the sample: Abell 1185 and Abell 2255, each at 99$\%$ confidence. In each case, the cluster had a larger fraction of radio galaxies than would be expected. For Abell 1185, these were exclusively lower luminosity radio galaxies (defined here as below $10^{23}$ W Hz$^{-1}$) while for Abell 2255 the more dramatic increase in radio galaxy fraction was for the higher radio luminosity galaxies. This result was expected based on the unusual number of extended radio galaxies noted in Section \ref{sec:results}. Abell 2255 also had a possible excess of low luminosity radio galaxies, with a signficance level of just over 87$\%$. When examining just the lower luminosity radio galaxies, Abell 2634 also appeared unusual in that it had a lower fraction of radio galaxies than would be expected (about 98$\%$ significance).

What types of galaxies are producing these results? To investigate this question, we examined the radio galaxy populations in light of their optical magnitudes. The same analysis was performed on subsets of the cluster galaxies corresponding to bright galaxies ($M_{R_c} \leq -22$), intermediate galaxies ($-21 \geq M_{R_c} > -22$), and faint galaxies ($-20 \geq M_{R_c} > -21$). As noted before, this yielded a larger number of inconclusive results due to low expected numbers of radio galaxies, but still provided useful insight for many of the clusters. It was confirmed (97.5$\%$) that the excess of high radio luminosity galaxies in Abell 2255 were the optically brightest galaxies \citep[which is to be expected, since the probability for a galaxy to have a powerful radio source increases with optical luminosity up to $\sim M^* - 0.5$;][]{ledl1996}. The excess in radio galaxy fraction in Abell 1185 corresponded to intermediate optical magnitude galaxies (97.1$\%$; similarly, Abell 539 appears to also have more of these type galaxies than the other clusters, at over 99$\%$). Dividing the sample into ranges in optical magnitude also revealed a second striking difference between Abell 2255 and the other clusters: Abell 2255 has a remarkable excess of optically-faint radio galaxies (over 99.9$\%$ significance). When performing the comparisons without the breakdown by optical magnitude, this population was missed because Abell 2255's radio fraction of intermediate magnitude galaxies was consistent with the rest of the sample.

\placefigure{fig-a2255rlf}

The results for Abell 2255 are depicted graphically in Figure \ref{fig-a2255rlf}. We have plotted radio galaxy fractions as a function of radio luminosity (i.e., a radio luminosity function) for Abell 2255 relative to the rest of the sample. The top left panel corresponds to all radio galaxies brighter than $M_{R_c}=-20$, where the increased radio galaxy fraction in Abell 2255 is evident. The other three panels correspond to the results for the three subsets created by optical magnitude.

It should be noted that the present discussion has not separated the radio galaxies into those powered by star formation and those powered by AGN, as was done in Section \ref{sec:over}. This separation is not possible for the entire comparison sample, as we lack spectroscopy for about half of the galaxies.\footnote{Note that velocities placing these galaxies in their respective clusters were obtained from the literature.} In general, the fainter galaxies are most often associated with star formation, the brighter galaxies with AGN (usually associated with elliptical host galaxies and lacking emission lines), and the intermediate galaxies representing a mixture of star forming galaxies and AGN (including emission line AGN such as Seyferts and LINERs). In the case of Abell 2255, there are 22 galaxies falling in the optically defined faint category (i.e., top right of Figure \ref{fig-a2255rlf}). The complete spectroscopy for the cluster enables us to note that 18 of these are star forming galaxies, with the remaining four including the two possible star forming galaxies with post-starburst spectra (J171207+640832 and J171400+640542) and a pair of weak emission line AGN. In the intermediate optical magnitude range, there are 16 galaxies evenly split between star forming and AGN. The eight bright galaxies are exclusively AGN, with all but one lacking emission lines. Thus, the excess of radio sources among the optically-faint galaxies is almost exclusively attributed to star forming galaxies in Abell 2255.

Since the radio galaxies are defined as cluster members only after they have been confirmed as such using spectroscopic measurements, the largest source of potential error in this calculation is the background correction. The effect of changing the assumed value of $\mathcal{N}$ is minimal, as it is applied to all clusters of the sample. The much more critical parameter is cosmic variance, or the large variation in galaxy counts over different regions of sky. To determine whether Abell 2255 only appears to have a high fraction of radio galaxies because it happens to lie in a region where the background is low, we investigated the radio galaxy fractions with different assumptions about the background correction. In these cases, we maintained $\mathcal{N}=1.26 \times 10^{-5}$ galaxies steradian$^{-1}$ for the other 19 clusters but lowered it for Abell 2255. To reduce the significance of Abell 2255's high net radio galaxy fraction to below 95$\%$ required that the background be $50\%$ lower. The result was even more striking for the optically faint galaxies. If {\it all} galaxies between $m_{R_c}^*+1$ and $m_{R_c}^*+2$ were cluster members (hence no background correction, and the lowest possible radio galaxy fraction), Abell 2255 would still exhibit an excess of radio galaxies in this magnitude interval at 97.5$\%$ confidence.

\section{Discussion}\label{sec:discuss}

The data presented indicate that Abell 2255 is an unusually active cluster of galaxies. It shows an excess of both powerful radio galaxies as well as optically faint star forming galaxies. The spectra of many of the star forming galaxies show evidence for current or recent bursts of star formation, strengthening the argument for increased activity in the cluster. Furthermore, the distribution of the star forming galaxies (and particularly those with evidence for current or recent starbursts) is elongated in a direction roughly perpendicular to the elongation of the X-ray emission and radio halo.

What is it about this cluster that is so conducive to these types of activity? We searched for correlations between the radio galaxy fractions and two simple parameters of the clusters. There was no evidence for any correlation of radio galaxy fraction with cluster richness (as parametrized by number of galaxies). Also, no correlation was found between radio galaxy fraction and the ``compactness'' of the clusters, described as the number of galaxies within 0.5 Mpc divided by the number of galaxies within 2 Mpc.

We are led to the notion that the dynamical state of Abell 2255 is somehow responsible for its unusually large number of radio galaxies. \citet{owen1999} found very different levels of activity for the clusters Abell 2125 and Abell 2645. These clusters were chosen as they have the same richness and lie at the same redshift, yet Abell 2125 has a large fraction of blue galaxies whereas Abell 2645 does not. This difference was confirmed via deep radio observations and optical spectroscopy, and the authors noted that Abell 2125 is a likely cluster-cluster merger whereas Abell 2645 appears more virialized. Consequently, they concluded that cluster-cluster mergers might have a dramatic effect on galaxies. Given the large amount of evidence that Abell 2255 is a cluster-cluster merger, we are led to the conclusion that such activity is probably responsible for the excess of radio galaxies in the cluster.

X-ray, radio, and optical data point toward the merger axis for Abell 2255 being aligned roughly East-West and close to the plane of the sky. This axis is easily seen in the X-ray emission of the cluster (see Figure \ref{fig-xray}), and corresponds well with the cluster's radio halo \citep{fere1997}. This is also revealed in optical images from the general alignment of the brighter galaxies at the center of the cluster. Simulations of a head-on collision between two clusters presented in \citet{burn1995} suggest that the merger axis is oriented $\approx30^\circ$ from the line of sight to the cluster, and that the system is viewed only a short time after the cores of the two progenitor clusters have crossed ($\approx0.13$ Gyr).

How can such a merger explain the types of radio galaxies observed and their distributions? The case for the powerful radio galaxies is fairly straightforward. Galaxies of this type are among the more massive ellipticals in clusters, and are consequently located near the centers of clusters. They should therefore be distributed along the merger axis, which is the general impression one gets from Figure \ref{fig-agndist}. Four of the FR I sources and the FR II source may be found approximately along the East-West axis and within $\sim1.5$ Mpc of the cluster center. 

However, while combining two clusters would clearly increase the overall number of powerful radio galaxies, an increased radio galaxy fraction for such sources would result only if previously quiescent elliptical galaxies were triggered into becoming powerful radio sources. Feretti et al. note this possibility in reference to J171329+640249, which has a radio morphology consisting of two compact yet strong lobes of emission (see Figure \ref{fig-double}). Galaxies with this type of morphology are typically much larger, leading them to suggest its small size was consistent with its radio emission being triggered by the recent merger. Additionally, mergers may prove necessary in the explanation of galaxies with head-tail morphologies. Such mergers can produce large relative velocities between the host galaxy and the intracluster gas. In the model of \citet{jone1979}, beams of radio-emitting material originate in the nucleus of the tailed radio galaxies. These beams are deflected by the host galaxy's ISM into the wake which follows behind the galaxy as it moves through the intracluster gas. Turbulence then serves to reaccelerate the particles in the wake and thereby power the radio emission in the tail. Alternatively, it is possible that no new powerful radio galaxies have been created in the merger. As a region rich in substructure (Paper 1), Abell 2255 may contain an excess of elliptical galaxies which were formerly associated with groups which have collected to form the cluster.

Cluster mergers may also initiate bursts of star formation in member gas-rich galaxies. \citet{gunn1972} noted that field galaxies falling into a cluster would experience ram pressure through interaction with the ICM. \citet{dres1983} argued that this could induce bursts of star formation as the molecular clouds of infalling galaxies get compressed. \citet{edge1990} related this to mergers and the Butcher-Oemler effect, since merger activity would have been more common in the past. In addition, the Butcher-Oemler effect is an averaged trend meaning that not all intermediate redshift clusters have increased blue fractions. This spread in blue fraction fits in with a spread in dynamical state.

However, the role of the ICM on the star formation of member galaxies is complicated by stripping. While denser molecular clouds may be compressed and lead to bursts of star formation, ram pressure appears to be highly effective at removing neutral gas from infalling galaxies. \citet{quil2000} argued that ram pressure stripping could remove the entire gas supply of the galaxy in a very short time period ($\sim10^8$ years). This is particularly true for less massive galaxies, such as the fainter star forming galaxies seen in Abell 2255. Once a cluster member galaxy has lost its gas reservoir, there is no reason to believe that it could resume a significant star forming episode.

\citet{roet1996} discussed the observational consequences of cluster-cluster mergers. Their simulations presented an attractive theory which explains both the nature of Abell 2255's star forming galaxies and their distribution. As a subcluster falls into a larger cluster, it develops a bow shock which serves to protect the subcluster galaxies and ICM. As the subcluster approaches the core of the primary cluster, this shock front slows and the subcluster galaxies pass through it. The large and rapid increase in ram pressure could initiate a burst of star formation, followed by truncation of the star formation due to gas stripping. While the subcluster galaxies are most affected by this process, it also applies to the galaxies in the primary cluster. The shock should also extend over Mpc scales before dissipating, thereby affecting many cluster galaxies. Based on observations of five nearby clusters with strong evidence for substructure, \citet{cald1997} argued for a similar evolutionary mechanism. The circumstances in Abell 2255 are nearly ideal should this be occurring. As it appears to be observed just after core passage, star forming galaxies in both the subcluster and primary cluster would have recently interacted with the shock front. The viewing geometry implies that star formation induced by such an effect would appear along a line centered on the core of the cluster and at right angles to the merger direction. Indeed, this is what is observed (see Figure \ref{fig-sfdist}).

As our straw model, we suggest that the peripheries of the progenitor clusters in the Abell 2255 merger system included a number of field-type galaxies. These galaxies would have been bound to the clusters but in fairly circular orbits which kept them well outside the destructive influence of their parent cluster's intracluster gas. The merger would subject them to the environmental influences discussed above, often initiating bursts of star formation. The evidence for these episodes respresenting starbursts rather than more normal star formation is seen in their spectral classifications. At least two are strong starbursts, and at least six are of the spectral type associated with dusty starbursts \citep{pogg1999,pogg2000}. Another galaxy has a k+a spectrum, indicating a burst of star formation has terminated in the galaxy within the past $\sim$1 Gyr. Given the effectiveness of ram pressure stripping, once the initiated starburst has run its course the galaxies will fade. The brighter spirals might end up as S0 galaxies, but the majority of the observed star forming galaxies are likely to become dwarf spheroidals.

The rapidity with which these galaxies burst and fade suggests that the combination of timing and observational limits is critical to their identification. This helps to explain the different results for the other cluster-cluster mergers of the sample. Abell 2256 is another excellent cluster-cluster merger candidate, but other than a potential excess of the most radio luminous galaxies it appears fairly normal. This cluster is believed to be an example of a pre-merger system \citep{sun2002}. Consequently, it is probably seen prior to the initiation of any bursts of star formation. The Coma cluster (Abell 1656) is another potential merger candidate. In fact, several authors have commented on the similarities between Coma and Abell 2255 \citep[e.g.,][]{davi1998}, including that each cluster has a radio halo and a pair of dominant galaxies. The evidence for Coma, however, suggests that it is being seen post-merger \citep{burn1994}. Any induced starbursts would have run their course, although the brighter ones might still be identified via post-starburst features. In fact, \citet{cald1993} and \citet{cald1997} do note that the Coma galaxies with post-starburst features are located primarily in the Southwest of the cluster, consistent with a past merger in which the starbursts occurred near core passage. A further aspect of the straw model is that imposing an optical magnitude cutoff of $M_{R_c}\leq-20$ means that any fainter galaxies which underwent bursts of star formation akin to those observed in Abell 2255 would have faded out of view. In fact, the deep spectroscopy in Abell 2255 actually detects star forming cluster galaxies {\it fainter} than $M_{R_c}=-20$ (see Table \ref{tbl-moreRG}). These include J170902+641728, a very compact faint galaxy ($M_{R_c}=-19.0$) and J170911+632940, a strong e(b) galaxy that lies just outside the 3 Mpc limit ($M_{R_c}=-19.5$). Another potential cluster starburst is J171346+641647, which has a velocity which formally places it outside of our cluster sample ($cz=28548$). It is also faint, but bright enough to have passed our magnitude cutoff ($M_{R_c}=-20.3$). The situation for Coma may be similar; the faint galaxy population is currently under study in a series of papers \citep[e.g.,][]{komi2001,moba2001,pogg2001}. They do find that the fainter Coma galaxies show more evidence for recent star formation.

It is also intriguing to note that Abell 1185, with its increased fraction of radio galaxies with intermediate optical magnitudes, is a cluster with possible signs of a merger. \citet{mahd1996} presented a fairly large number of velocities for Abell 1185 galaxies, and noted significant evidence for substructure found from these velocities. In addition, a comparison with the Einstein X-ray data indicates that the centroid of the X-ray emission is offset from the optically-brightest galaxy by 3\arcmin. However, spectroscopy of these galaxies reveals that many are AGN \citep{mill2002}.

Another attractive feature of this explanation is its applicability to higher redshift clusters. Should mergers be important to cluster galaxy evolution, we would expect clusters to show increased signs of activity at earlier epochs when clusters were being assembled \citep[e.g.,][]{edge1990}. It might also reconcile the results found by the MORPHS and CNOC collaborations for intermediate clusters. The MORPHS clusters were generally selected optically, while the CNOC clusters are X-ray selected. Consequently, the MORPHS clusters are more likely to contain multiple bright galaxies (and hence, frequently cluster-cluster mergers) whereas the CNOC clusters are more likely to be virialized systems with bright central X-ray peaks. The populations of the two samples would thereby differ, with the MORPHS clusters showing large numbers of starburst and post-starburst systems and the CNOC clusters showing more quiescent histories for star forming galaxies. This explanation has been noted previously \citep[e.g.,][]{elli2001}.

Of course, the application of a single evolutionary model to the cluster is an over-simplification. Not all of the radio sources fit in with the above picture. There are several galaxies with spectra representative of fairly normal star formation located at various points along the periphery of the cluster. These are not aligned with the main distribution of star forming galaxies, nor do they show any evidence for unsual bursts of star formation activity. They do not therefore require any exotic explanation, as they could simply be normal star forming galaxies just entering the cluster. The powerful radio galaxies also suggest that the cluster's activity is not confined to the merger of only two partners. While four of the FR-type sources are located in the center of the system and aligned roughly along the merger axis, a fifth is a full 1.2 Mpc to the East from the cluster center along this axis (J171509+640254). Another lies $\sim1.3$ Mpc to the South (J171316+634738), while the most extreme example is just outside 3 Mpc to the Northeast (J171530+643952). It would seem likely that these radio sources were associated with groups other than the two primary merger partners. Their presence is still difficult to explain given the lack of any obvious associated X-ray emission in their vicinities (Figure \ref{fig-xray}, \ref{fig-agndist}).

\section{Conclusions}\label{sec:conclude}

Abell 2255 is an unusually active cluster of galaxies. Through our 1.4 GHz radio observations, which represent the most sensitive wide-field investigation of the cluster ever undertaken, we have identified 46 cluster radio galaxies. Included among these are six sources with FR I or FR II radio morphologies, which represents a marked excess of such sources over typical clusters. A more striking excess is observed in the faint star forming population. This result is robust, being insensitive to the assumed background correction and cosmic variance. Thus, galaxies in Abell 2255 are more likely to be radio sources than comparable galaxies in other clusters.

The reason for this enhancement appears to be a fortuitous combination of factors related to the merger stage, geometry, and our observational limits. Abell 2255 is an excellent example of a dynamically-active cluster, viewed nearly at a right angle to a merger axis and shortly after the cores of the progenitor clusters have crossed. Models suggest that this situation implies that a large number of galaxies have recently crossed the shock front that formed between the progenitors, thereby causing a large spike in pressure on their interstellar matter and promoting a burst of star formation. The viewing geometry insures that these galaxies appear distributed roughly along a line perpendicular to the merger axis, an effect which is seen in the data. Furthermore, the spectra of the galaxies often indicate a burst of star formation rather than simple truncation. Interaction with the ICM will presumably remove the reservoirs of neutral gas that these galaxies possess, so once they have exhausted their molecular gas in their current star formation episodes they will fade into a quiescent population. Hence, should the same cluster be viewed roughly one Gyr hence it would likely appear fairly normal in regards to its radio population.

This model clearly requires additional testing. In addition to further numerical simulations, it needs to be evaluated using a larger sample of cluster-cluster mergers. These necessarily include merger systems viewed at a range of stages (from pre-core passage through post-core passage) and viewing geometries, plus deeper observational limits to better probe the faint galaxy population which appears critical to our understanding. This latter point is particularly true at higher redshift, where merger activity is likely more common but absolute observational limits are higher. Of course, the evaluation of factors such as merger stage and viewing geometry is difficult and undoubtedly requires both increased data (X-ray, galaxy kinematics, etc.) and numerical modeling of mergers.

\acknowledgments

The authors thank Bill Oegerle, John Hill, and Rajib Ganguly for access to their spectroscopic database of Abell 2255, and an anonymous referee for insightful comments. NAM thanks the National Radio Astronomy Observatory for a predoctoral fellowship, and acknowledges the support of a National Research Council Associateship Award at NASA's Goddard Space Flight Center.

Funding for the creation and distribution of the SDSS Archive has been provided by the Alfred P. Sloan Foundation, the Participating Institutions, the National Aeronautics and Space Administration, the National Science Foundation, the U.S. Department of Energy, the Japanese Monbukagakusho, and the Max Planck Society. The SDSS Web site is {\ttfamily http://www.sdss.org/}.

The Participating Institutions are The University of Chicago, Fermilab, the Institute for Advanced Study, the Japan Participation Group, The Johns Hopkins University, the Max-Planck-Institute for Astronomy (MPIA), New Mexico State University, Princeton University, the United States Naval Observatory, and the University of Washington.

\clearpage

\begin{deluxetable}{l l r r r r r c l l}
\tablecolumns{10}
\tablecaption{Cluster Radio Galaxies\label{tbl-clusRG}}
\tablewidth{0pt}
\tablenum{1}
\tablehead{
\colhead{RA} & \colhead{Dec} & \colhead{$cz$} & 
\colhead{$\delta cz$} & \colhead{$m_{R_c}$} & 
\colhead{$S_{1.4}$} & \colhead{$\delta S_{1.4}$} & \colhead{Sep} &
\colhead{Class} & \colhead{MORPHS} \\
\multicolumn{2}{c}{(J2000)} & 
\multicolumn{2}{c}{[km s$^-1$]} & \colhead{} & 
\multicolumn{2}{c}{[mJy]} & \colhead{[\arcsec]} & \colhead{} & \colhead{}
}
\startdata
17:08:55.7 & 63:41:55 & 27409 & 42 & 17.44 &    0.400 & 0.093 & 0.9 & SF & e(c) \\
17:10:31.5 & 64:39:15 & 23661 & 17 & 15.80 &    3.000 & 0.105 & 0.4 & SF & e(c) \\
17:10:36.2 & 64:20:03 & 23347 & 18 & 16.93 &    2.211 & 0.089 & 0.6 & AGNl & e(n) \\
17:10:51.2 & 63:51:13 & 24884 & 17 & 16.15 &    0.265 & 0.063 & 1.5 & AGNl & e(c) \\
17:10:52.5 & 63:39:17 & 24200 & 22 & 16.25 &    0.330 & 0.099 & 0.6 & AGNl & e(n) \\
17:10:57.4 & 63:51:59 & 24864 & 14 & 16.93 &    0.331 & 0.083 & 1.4 & SF & e(c) \\
17:11:06.6 & 63:52:00 & 24850 & 41 & 16.36 &    0.313 & 0.068 & 0.1 & AGNo & k \\
17:11:17.1 & 64:19:20 & 23021 & 20 & 16.61 &    1.060 & 0.086 & 0.7 & SF & e(c) \\
17:11:17.4 & 64:20:33 & 23050 & 28 & 16.67 &    0.582 & 0.133 & 0.7 & SF & e(c)\tablenotemark{a} \\
17:11:29.0 & 63:58:49 & 21365 & 11 & 16.20 &    0.671 & 0.064 & 0.8 & SF & e(c) \\
17:11:34.5 & 63:53:38 & 22974 & 14 & 16.96 &    0.326 & 0.054 & 1.0 & SF & e(a) \\
17:11:39.6 & 64:10:07 & 22963 & 13 & 17.19 &    1.177 & 0.069 & 0.9 & SF & e(a) \\
17:11:45.1 & 64:12:16 & 24614 & 14 & 17.32 &    0.338 & 0.066 & 1.0 & SF & e(a) \\
17:11:46.0 & 64:21:32 & 24593 & 16 & 15.43 &    1.939 & 0.089 & 0.3 & AGNo & k \\
17:11:57.7 & 64:03:21 & 25065 & 25 & 16.36 &    0.655 & 0.073 & 1.3 & AGNo & k \\
17:11:59.3 & 64:16:36 & 24446 & 16 & 16.80 &    0.493 & 0.079 & 1.2 & SF & e(c) \\
17:12:06.9 & 64:08:32 & 24681 & 26 & 17.02 &    0.475 & 0.062 & 0.7 & AGNo\tablenotemark{b} & k+a \\
17:12:16.2 & 64:02:13 & 21416 & 21 & 15.46 &   12.6\phn\phn & 0.3\phn\phn & \tablenotemark{c} & AGNo & k \\
17:12:22.9 & 64:08:29 & 22551 & 14 & 16.88 &    1.277 & 0.081 & 0.9 & SF & e(c) \\
17:12:23.2 & 64:01:57 & 23999 & 42 & 15.52 &  287.7\phn\phn & 0.5\phn\phn & \tablenotemark{c} & AGNo & k \\
17:12:28.7 & 64:06:33 & 22227 & 17 & 17.28 &    0.345 & 0.068 & 0.5 & SF & e(c) \\
17:12:34.1 & 64:05:50 & 22838 & 19 & 17.38 &    0.817 & 0.078 & 1.4 & SF & e(b) \\
17:12:43.0 & 63:48:35 & 22472 & 14 & 16.49 &    0.653 & 0.086 & 0.3 & SF & e(a) \\
17:12:48.4 & 63:42:58 & 22541 & 20 & 16.05 &    2.768 & 0.066 & 0.3 & AGNs & e(n) \\
17:12:51.2 & 64:04:24 & 22181 & 24 & 15.98 &    0.643 & 0.097 & 1.0 & AGNs & e(n) \\
17:13:02.0 & 64:03:09 & 24623 & 18 & 15.03 &    1.430 & 0.073 & 1.0 & AGNo & k \\
17:13:02.4 & 64:10:09 & 22959 & 15 & 16.34 &    0.443 & 0.062 & 0.4 & SF & e(a) \\
17:13:03.8 & 64:07:02 & 24313 & 19 & 15.46 &   65.9\phn\phn & 0.4\phn\phn & \tablenotemark{c} & AGNo & k \\
17:13:15.5 & 64:24:18 & 24965 & 18 & 16.87 &    0.237 & 0.081 & 1.5 & SF & e(c) \\
17:13:16.0 & 63:47:38 & 24843 & 15 & 14.88 &  142.2\phn\phn & 0.6\phn\phn & \tablenotemark{c} & AGNo & k \\
17:13:19.9 & 63:54:28 & 27090 & 22 & 16.42 &    0.830 & 0.095 & 4.6 & SF & e(c) \\
17:13:20.5 & 64:00:34 & 23954 & 18 & 17.38 &    0.535 & 0.071 & 2.8 & SF & e(b) \\
17:13:29.1 & 64:02:49 & 23483 & 19 & 15.02 &  248.3\phn\phn & 0.3\phn\phn & \tablenotemark{c} & AGNo & k \\
17:13:33.9 & 63:49:26 & 25826 & 18 & 15.56 &    0.817 & 0.080 & 0.1 & SF & k+a \\
17:13:36.9 & 63:59:19 & 21239 & 17 & 16.60 &    0.357 & 0.053 & 0.9 & AGNs & e(n) \\
17:13:43.0 & 64:04:55 & 25423 & 26 & 15.64 &    2.766 & 0.074 & 0.6 & AGNo & k \\
17:13:43.3 & 63:50:06 & 22509 & 27 & 17.34 &    0.661 & 0.105 & 1.2 & SF & e(c)\tablenotemark{d} \\
17:13:51.2 & 64:00:13 & 26290 & 15 & 16.10 &    1.432 & 0.149 & 1.2 & SF & e(c) \\
17:13:52.0 & 64:07:10 & 24843 & 13 & 17.32 &    0.365 & 0.097 & 1.8 & SF & e(c) \\
17:14:00.4 & 64:05:42 & 26294 & 78 & 17.11 &    0.369 & 0.095 & 0.4 & AGNo & k\tablenotemark{e} \\
17:14:24.3 & 63:39:38 & 24388 & 17 & 16.92 &    1.005 & 0.092 & 0.8 & SF & e(c) \\
17:14:47.4 & 64:35:41 & 24035 & 20 & 16.52 &    1.001 & 0.112 & 0.9 & SF & e(c) \\
17:15:02.3 & 64:35:41 & 23763 & 19 & 16.76 &    0.511 & 0.127 & 2.1 & SF & e(c) \\
17:15:09.1 & 64:02:54 & 24000 & 25 & 15.02 &   90.3\phn\phn & 0.8\phn\phn & \tablenotemark{c} & AGNo & k \\
17:17:10.4 & 64:19:38 & 26229 & 22 & 15.32 &    0.463 & 0.124 & 2.9 & AGNl & k \\
17:18:01.2 & 64:22:54 & 24179 & 40 & 16.60 &    0.935 & 0.140 & 2.7 & SF & e(c) \\
\enddata

\tablenotetext{a}{H$\delta$ absorption is sufficient for classification as an e(a) galaxy, but measurement is noisy and we tentatively classify the galaxy e(c).}
\tablenotetext{b}{The lack of optical emission lines formally places this galaxy in the AGN category. However, there is evidence that these galaxies are heavily dust-obscured star forming galaxies (see text).}
\tablenotetext{c}{See Figure \ref{fig-otrg}--\ref{fig-embryo}.}
\tablenotetext{d}{Wavelength coverage of spectrum misses \OII; strong emission lines suggest this is a potential e(b) galaxy.}
\tablenotetext{e}{Potential k+a galaxy, although spectrum is relatively noisy.}

\tablecomments{All velocities and associated errors have been taken from Paper 1, which includes SDSS EDR data. The $R_c$ magnitudes were calculated for the Gunn-Oke aperture, and have an associated error of $\lesssim0.05$ mags. The classes are determined from the galaxy spectra, and are defined as follows: SF -- star forming galaxy; AGNl -- AGN with optical spectrum dominated by an old stellar population but with weak emission of \NII{} and sometimes \SII; AGNo -- AGN with pure absorption line spectrum representative of old stellar population stars; AGNs -- AGN with emission line spectrum (Seyfert).}

\end{deluxetable}

\begin{deluxetable}{l l r r r r r c l l}
\tablecolumns{10}
\tablecaption{Additional Radio Galaxies\label{tbl-moreRG}}
\tabletypesize{\small}
\tablewidth{438pt}
\tablenum{2}
\tablehead{
\colhead{RA} & \colhead{Dec} & \colhead{$cz$} & 
\colhead{$\delta cz$} & \colhead{$m_{R_c}$} & 
\colhead{$S_{1.4}$} & \colhead{$\delta S_{1.4}$} & \colhead{Sep} &
\colhead{Class} & \colhead{Notes} \\
\multicolumn{2}{c}{(J2000)} & 
\multicolumn{2}{c}{[km s$^-1$]} & \colhead{} & 
\multicolumn{2}{c}{[mJy]} & \colhead{[\arcsec]} & \colhead{} & \colhead{}
}
\startdata
17:06:22.7 & 64:03:59 &  92908 &  54 & 18.12 &    0.341 & 0.079 & 3.5 & SF   & \tablenotemark{a} \\
17:06:34.0 & 63:45:09 &  43075 &  47 & 17.35 &    0.765 & 0.150 & 1.7 & SF   & \tablenotemark{a} \\
17:08:00.6 & 63:39:53 &  52922 &  28 & 17.76 &    6.447 & 0.097 & 0.6 & AGNs & \tablenotemark{a} \\
17:08:10.4 & 63:37:52 &   8146\tablenotemark{b} &  22 & 11.97 &    8.5\phn\phn & 0.8\phn\phn & --  & Pair & \tablenotemark{c,d} \\
17:08:37.8 & 64:01:45 & 136405\tablenotemark{e} &  57 & 18.81 &    0.379 & 0.086 & 0.8 & AGNo & \tablenotemark{a} \\
17:08:45.1 & 63:47:46 &  16257 &  19 & 15.71 &    0.899 & 0.119 & 0.3 & SF   & \tablenotemark{c} \\
17:09:02.1 & 64:17:28 &  26803\tablenotemark{e} & 218 & 18.53 &    0.787 & 0.090 & 0.5 & SF?  & \tablenotemark{f,g} \\
17:09:11.1 & 63:29:40 &  23847 &  20 & 18.04 &    0.441 & 0.108 & 0.7 & SF   & \tablenotemark{d,f,g} \\
17:09:11.1 & 64:22:10 & 610976\tablenotemark{e} & 316 & ---   &    0.784 & 0.096 & 0.6 & QSO  & \tablenotemark{a} \\
17:09:34.8 & 63:59:57 &  70975 &  61 & 17.47 &    0.366 & 0.073 & 1.7 & AGN? & \tablenotemark{a} \\
17:09:41.5 & 63:52:02 &  58861 &  26 & 17.75 &    0.267 & 0.064 & 1.9 & SF   & \tablenotemark{a,g} \\
17:09:56.1 & 64:41:26 &  56001 & 120 & 17.98 &    0.874 & 0.194 & 3.7 & AGNo & \tablenotemark{a,g} \\
17:10:09.8 & 64:22:57 &  76841 &  39 & 17.30 &    0.611 & 0.095 & 0.3 & SF   & \tablenotemark{a} \\
17:10:10.3 & 64:11:47 &  51410 &  27 & 19.18 &    0.269 & 0.070 & 2.1 & SF   & \tablenotemark{a} \\
17:10:25.5 & 63:17:44 &  52375 &  55 & 17.18 &    1.415 & 0.142 & 2.0 & AGNo & \tablenotemark{a} \\
17:10:44.5 & 64:17:02 &  71073 &  14 & 17.15 &    0.718 & 0.098 & 0.6 & SF   & \tablenotemark{a} \\
17:11:01.7 & 64:01:34 &  28652 &  39 & 15.72 &    0.196 & 0.056 & 0.5 & AGNo & \tablenotemark{a} \\
17:11:04.0 & 63:42:43 &  24965 &  23 & 17.64 &    0.435 & 0.090 & 0.5 & SF   & \tablenotemark{f,g} \\
17:11:17.9 & 64:39:02 &  67657 &  20 & 18.32 &    0.256 & 0.073 & 1.0 & SF   & \tablenotemark{a} \\
17:11:36.5 & 63:44:45 &  45104 &  26 & 17.71 &    0.218 & 0.078 & 0.7 & SF   & \tablenotemark{a,g} \\
17:12:23.0 & 63:49:06 &  45183 &  16 & 17.39 &    0.432 & 0.062 & 0.4 & SF   & \tablenotemark{a} \\
17:13:07.4 & 64:05:01 & 137005\tablenotemark{e} &  67 & 19.26 &    0.422 & 0.125 & 4.6 & AGNo & \tablenotemark{a} \\
17:13:19.7 & 63:42:17 &  45163 &  15 & 17.12 &    0.304 & 0.095 & 1.3 & SF   & \tablenotemark{a} \\
17:13:30.2 & 64:42:53 & 314182\tablenotemark{e} & 196 & 17.65 &    0.915 & 0.081 & 0.6 & QSO  & \tablenotemark{a} \\
17:13:45.5 & 64:14:47 &  28558 &  15 & 17.16 &    0.586 & 0.080 & 0.4 & SF   & \tablenotemark{a} \\
17:13:59.5 & 64:09:40 & 408616\tablenotemark{e} & 342 & 18.25 &    0.955 & 0.065 & 0.2 & QSO  & \tablenotemark{a} \\
17:14:05.6 & 64:16:02 &  74007 &  40 & 17.44 &    5.900 & 0.089 & 1.9 & AGNo & \tablenotemark{a,h} \\
17:14:18.8 & 64:38:09 &  16062 &  20 & 15.57 &    1.131 & 0.103 & 0.4 & SF   & \tablenotemark{c} \\
17:14:32.8 & 63:39:29 & 107925\tablenotemark{e} &  54 & 18.45 &    0.800 & 0.090 & 0.8 & AGN? & \tablenotemark{a} \\
17:14:37.1 & 63:59:38 &   9382 &  62 & 17.74 &    0.231 & 0.086 & 1.1 & AGN? & \tablenotemark{c,g} \\
17:15:01.0 & 64:28:51 &  31000 &  19 & 17.09 &    0.824 & 0.098 & 1.1 & SF   & \tablenotemark{a} \\
17:15:29.5 & 63:44:53 &  53081 &  31 & 17.32 &    0.353 & 0.086 & 1.4 & SF   & \tablenotemark{a} \\
17:15:30.0 & 64:39:52 &  23624\tablenotemark{e,i} &  92 & 15.00 &  116.3\phn\phn & 2.3\phn\phn & --  & AGNo & \tablenotemark{d,f,j} \\
17:17:23.8 & 64:17:24 &  10322 &  25 & 14.90 &    2.164 & 0.108 & 0.4 & AGNl & \tablenotemark{c} \\
17:18:14.5 & 64:17:36 &  31016 &  25 & 17.08 &    2.329 & 0.108 & 0.4 & AGNl & \tablenotemark{a} \\
17:18:31.1 & 64:25:05 &  26746 &  48 & 15.51 &    0.861 & 0.159 & 3.0 & AGNl & \tablenotemark{d,f} \\
\enddata

\tablenotetext{a}{Background galaxy.}
\tablenotetext{b}{Velocity and error from \citet{stra1992}. Galaxy pair (UGC10731 or VV726), where each galaxy is a radio source but the flux measurement and velocity correspond to the system.}
\tablenotetext{c}{Foreground galaxy.}
\tablenotetext{d}{Outside the 3 Mpc radial limit of the study.}
\tablenotetext{e}{Velocity and error taken from the SDSS EDR.}
\tablenotetext{f}{Cluster member.}
\tablenotetext{g}{Optical magnitude too faint to be formally included in the radio galaxy study.}
\tablenotetext{h}{See Figure \ref{fig-faker}.}
\tablenotetext{i}{The SDSS velocity for this galaxy was flagged due to inconsistent measurements by their automated routines. Inspection of the spectrum and a prior velocity measurement from NED clearly indicate it is consistent with cluster membership.}
\tablenotetext{j}{See Figure \ref{fig-bean}.}

\tablecomments{Velocities and associated errors have been taken from Paper 1, unless otherwise noted. The $R_c$ magnitudes were calculated for the Gunn-Oke aperture, and have an associated error of $\lesssim0.05$ mags. The classes are determined from the galaxy spectra, and are defined as follows: SF -- star forming galaxy; AGNl -- AGN with optical spectrum dominated by an old stellar population but with weak emission of \NII{} and sometimes \SII; AGNo -- AGN with pure absorption line spectrum representative of old stellar population stars; AGNs -- AGN with emission line spectrum (Seyfert); QSO -- quasar designation from SDSS EDR. Note that J170902+641728, J170911+632940, J171010+642257, J171118+643902, and J171346+641447 are compact starbursts misidentified as QSOs by the automated routines of the SDSS EDR. False identifications of this type are among the principal contaminants of the SDSS QSO sample \citep{rich2002}.}

\end{deluxetable}

\begin{deluxetable}{r r r r r r r c r r r r r r}
\tablecolumns{14}
\tablecaption{Galaxy Counts\label{tbl-rgnums}}
\tabletypesize{\small}
\tablewidth{469pt}
\tablenum{3}
\tablehead{
\colhead{} & \multicolumn{6}{c}{2 Mpc Sample} & 
\colhead{} & \multicolumn{6}{c}{3 Mpc Sample}\\

\cline{2-7} \cline{9-14}\\

\colhead{} & \multicolumn{2}{c}{Faint} & \multicolumn{2}{c}{Intermediate} & \multicolumn{2}{c}{Bright} & \colhead{} &
\multicolumn{2}{c}{Faint} & \multicolumn{2}{c}{Intermediate} & \multicolumn{2}{c}{Bright}\\

\colhead{Cluster} & \colhead{$N_{RG}$} & \colhead{$N_{tot}$} &
\colhead{$N_{RG}$} & \colhead{$N_{tot}$} & \colhead{$N_{RG}$} & \colhead{$N_{tot}$} & \colhead{} &
\colhead{$N_{RG}$} & \colhead{$N_{tot}$} & \colhead{$N_{RG}$} & \colhead{$N_{tot}$} & \colhead{$N_{RG}$} & \colhead{$N_{tot}$}
}
\startdata
A262 & 0 & 59.1 & 3 & 23.5 & 4 & 12.2 & & 0 & 87.8 & 4 & 28.4 & 4 & 13.1 \\
A347 & 1 & 53.2 & 6 & 38.3 & 3 & 16.1 & & 1 & 76.7 & 7 & 66.8 & 4 & 24.0 \\
A397 & 1 & 36.6 & 4 & 22.2 & 0 & 2.0 & & 2 & 35.3 & 5 & 23.8 & 1 & 0.6 \\
A400 & 2 & 46.1 & 4 & 16.9 & 0 & 2.7 & & 2 & 41.1 & 4 & 13.9 & 1 & 1.0 \\
A426 & 1 & 105.4 & 8 & 102.5 & 7 & 21.9 & & 1 & 131.4 & 11 & 148.3 & 9 & 26.4 \\
A539 & 0 & 53.2 & 9 & 20.5 & 3 & 8.2 & & 1 & 57.9 & 11 & 60.7 & 4 & 6.9 \\
A569 & 1 & 48.9 & 0 & 16.7 & 2 & 1.9 & & 2 & 87.6 & 0 & 33.6 & 6 & 6.5 \\
A634 & 1 & 34.3 & 4 & 15.3 & 1 & 12.8 & & 1 & 42.0 & 4 & 14.7 & 1 & 11.2 \\
A779 & 0 & 31.2 & 2 & 5.1 & 0 & 1.0 & & 0 & 32.7 & 2 & 3.3 & 0 & 0.8 \\
A1185 & 0 & 23.8 & 8 & 16.7 & 2 & 2.5 & & 1 & 40.3 & 10 & 53.3 & 4 & 3.7 \\
A1267 & 1 & 33.5 & 3 & 9.6 & 0 & 0.1 & & 2 & 53.5 & 5 & 13.1 & 0 & 0.1 \\
A1367 & 2 & 66.8 & 7 & 43.2 & 3 & 12.1 & & 2 & 111.8 & 9 & 56.7 & 3 & 15.9 \\
A1656 & 5 & 157.6 & 7 & 78.9 & 3 & 15.0 & & 7 & 194.1 & 9 & 93.0 & 6 & 19.7 \\
A2162 & 0 & 51.0 & 1 & 13.5 & 1 & 2.5 & & 0 & 96.5 & 2 & 21.8 & 2 & 4.6 \\
A2197 & 1 & 61.9 & 6 & 48.0 & 5 & 16.6 & & 4 & 124.8 & 11 & 75.9 & 7 & 29.9 \\
A2199 & 2 & 92.4 & 10 & 43.1 & 2 & 18.7 & & 5 & 146.8 & 13 & 71.2 & 7 & 29.0 \\
A2255 & 7 & 97.3 & 10 & 65.3 & 7 & 12.4 & & 10 & 84.9 & 11 & 66.1 & 7 & 9.6 \\
A2256 & 1 & 105.8 & 10 & 72.7 & 6 & 8.6 & & 1 & 101.0 & 10 & 81.8 & 7 & 7.2 \\
A2634 & 2 & 131.2 & 4 & 47.3 & 3 & 22.4 & & 5 & 178.8 & 5 & 60.4 & 3 & 30.4 \\
A2666 & 0 & 18.4 & 1 & 11.1 & 0 & 0.7 & & 4 & 41.6 & 2 & 22.2 & 2 & 3.0 \\
\enddata

\tablecomments{Galaxy counts for the twenty Abell clusters of the composite sample. Results are presented for galaxies within both 2 Mpc and 3 Mpc of the cluster centers. $N_{RG}$ refers to the number of radio galaxies and $N_{tot}$ refers to the total number of background-corrected galaxies, each within the specified optical magnitude range. The ranges correspond to: Faint -- $-20\geq M_R > -21$, Intermediate -- $-21\geq M_R > -22$, and Bright -- $M_R \leq -22$. Although this table does not divide the radio detections into high and low luminosity classes based on a cutoff of $10^{23}$ W Hz$^{-1}$, it should be noted that the high radio luminosity sources are almost exclusively the optically brightest galaxies (no faint galaxies are associated with high radio luminosities, and only seven intermediate galaxies are).}
\end{deluxetable}

\clearpage

\begin{figure}
\figurenum{1}
\caption{Color-magnitude plot for Abell 2255, based on SDSS data. The filled circles represent galaxies whose SDSS velocities confirm they are cluster members, with open triangles and circles representing foreground and background galaxies, respectively. Overplotted crosses indicate the cluster radio galaxies (which include several sources without SDSS velocities). The large six-pointed crosses represent the pair of radio galaxies deemed to be background sources based on this color-magnitude diagram (see text).\label{fig-colors}}
\end{figure}

\begin{figure}
\figurenum{2}
\caption{Overlays of radio contours on $R_c$ images for galaxies with FR morphologies. All of the radio images were made using a 5.9\arcsec{} restoring beam, and the base contour level is specified for each figure. The contours are made at -2, 2, 3, 5, 8, 13, 21, and 34 times the base value. At left is ``The Original TRG,'' and to the right is The Sidekick \citep[quoted source names are taken from ][]{harr1980}. The rms sensitivity for this image is 31 $\mu$Jy/beam.\label{fig-otrg}}
\end{figure}

\begin{figure}
\figurenum{3}
\caption{``The Double,'' with an rms sensitivity of 43 $\mu$Jy/beam.\label{fig-double}}
\end{figure}

\begin{figure}
\figurenum{4}
\caption{``The Goldfish,'' with an rms sensitivity of 34 $\mu$Jy/beam.\label{fig-goldfish}}
\end{figure}

\begin{figure}
\figurenum{5}
\caption{``The Beaver,'' with an rms sensitivity of 37 $\mu$Jy/beam.\label{fig-beaver}}
\end{figure}

\begin{figure}
\figurenum{6}
\caption{``The Embryo,'' with an rms sensitivity of 42 $\mu$Jy/beam.\label{fig-embryo}}
\end{figure}

\begin{figure}
\figurenum{7}
\caption{``The Bean,'' with an rms sensitivity of 134 $\mu$Jy/beam. This galaxy lies over 3 Mpc from the cluster center, and consquently is outside the formal limit for our study.\label{fig-bean}}
\end{figure}

\begin{figure}
\figurenum{8}
\caption{This galaxy is a background source at $z\approx0.25$. The rms level of the map is 44 $\mu$Jy/beam.\label{fig-faker}}
\end{figure}

\begin{figure}
\figurenum{9}
\caption{ROSAT PSPC observation of Abell 2255 for 0.5 -- 2.0 keV band. The map has been smoothed by a 45\arcsec{} Gaussian. The large circle indicates the 3 Mpc radial limit of our investigation.\label{fig-xray}}
\end{figure}

\begin{figure}
\figurenum{10}
\caption{The distribution of AGN in Abell 2255. The diamonds signify galaxies whose spectra are representative of old stellar populations, while the triangles are emission line AGN. An additional cross has been overplotted on the k+a galaxy without any optical emission lines. Note that these may be powered by star formation and not an AGN (see text). The large circle indicates the 3 Mpc radial limit of our investigation.\label{fig-agndist}}
\end{figure}

\begin{figure}
\figurenum{11}
\caption{The distribution of star forming galaxies in Abell 2255. The five-pointed stars signify normal galaxies (class e(c)), the many-pointed stars are starbursts (class e(b)), the triangles are dusty starbursts (class e(a)), the circle is the galaxy with a post-starburst spectrum (class k+a) but with H$\alpha$ emission, and the cross is the galaxy with post-starburst spectrum and no detected optical emission lines.\label{fig-sfdist}}
\end{figure}

\begin{figure}
\figurenum{12}
\caption{Starburst galaxy. Note disturbed appearance, with ``streamers'' (identified by arrows) pointing to the Southeast. The radio emission is also offset in this direction. The restoring beam is 5.9\arcsec{}, and the rms sensitivity is 39 $\mu$Jy/beam.\label{fig-ebgal}}
\end{figure}

\begin{figure}
\figurenum{13}
\caption{The cumulative radio luminosity function for Abell 2255, expressed as the fraction of galaxies hosting a radio source at the indicated radio luminosity or greater. Solid lines and filled circles represent the composite data from nineteen other nearby Abell clusters, while the dotted line and open triangles represent Abell 2255 (offset slightly in {\it x} for clarity). For this determination, only galaxies within 2 Mpc of the cluster centers were used. At top left is the RLF for all galaxies in the sample ($M_R\leq-20$), while the remaining three plots are based on cuts in optical magnitude. Note that the apparent flattening of the RLFs below $\log (L_{1.4GHz})\sim21.8$ is due to incompleteness.\label{fig-a2255rlf}}
\end{figure}

\end{document}